\pdfoutput=1  

\documentclass[10pt,a4paper]{article}

\usepackage[utf8]{inputenc}
\usepackage[T1]{fontenc}
\usepackage{lmodern}
\usepackage{microtype}
\usepackage{geometry}
\usepackage{hyperref}
\usepackage{amsmath, amssymb}
\usepackage{graphicx}
\usepackage{authblk} 

\hypersetup{
	colorlinks = true,
	urlcolor   = blue,
	linkcolor  = blue,
	citecolor  = blue,
	pdftitle   = {Your Title},
	pdfauthor  = {Author Names}
}

\title{Decomposition of the total wave aberration in generalized optical systems}

\author[1,*]{Mateusz Oleszko}
\author[1,2]{Ralf Hambach}
\author[1,2]{Herbert Gross}

\affil[1]{Friedrich-Schiller-Universit\"at Jena, Institute of Applied Physics, 07745 Jena, Germany}
\affil[2]{Fraunhofer Institute of Applied Optics and Precision Engineering, 07745 Jena, Germany}
\affil[*]{Corresponding author: mat.oleszko@gmail.com}

\date{%
	\vspace{0.5em}
	\small DOI: \href{https://doi.org/10.1364/JOSAA.34.001856}{10.1364/JOSAA.34.001856} \\
	\vspace{0.8em}
	{Uploaded: }\today \\ {First published: September 19, 2017}
}

\begin{document}
	
	\maketitle
	
	\begin{abstract}
		The increasing use of freeform optical surfaces raises the demand for optical design tools developed for generalized systems. In the design process surface-by-surface aberration contributions are of special interest. The expansion of the wave aberration function into field and pupil dependent coefficients is an analytical method used for that purpose. An alternative numerical approach utilizing data from the trace of multiple ray sets is proposed. The optical system is divided into segments of the optical path measured along the chief ray. Each segment covers one surface and the distance to the subsequent surface. Surface contributions represent the change of the wavefront that occurs due to propagation through individual segments.  Further, the surface contributions are divided with respect to their phenomenological origin into intrinsic, induced and transfer components. Each component is determined from a separate set of rays. The proposed method does not place any constraints on the system geometry or the aperture shape. However, here we concentrate on near-circular apertures and specify the resulting wavefront error maps using an expansion into Zernike polynomials. Hence, for the first time additive surface Zernike contributions are obtained.
	\end{abstract}
	
	\section{Introduction}
	\label{sec:1}
	The expansion of the wave aberration function into field and pupil dependent coefficients is an analytical tool for investigating surface contributions to the total wave aberration in axially symmetric systems. The fourth order aberration coefficients are given in terms of the famous Seidel sums. Currently, the expansion up to the sixth order is provided in literature \cite{1}. These sixth order coefficients assigned to each surface are further divided due to their origin into intrinsic and extrinsic components. The introduction of nodal aberration theory \cite{2} expands the functionality of the approach for the design of symmetry free optical systems \cite{3}.  Aberration coefficients are nevertheless often difficult to interpret due to their large number and complex dependencies on field and pupil. Hence, to better understand the results of the aberration function expansion, its coefficients are associated with Zernike Fringe polynomial set \cite{4, 5, 6}. Another approach, developed for systems with no symmetry constraints, is to consider the surface aberration contributions in the phase space \cite{7}.\\
	Wave aberration of optical systems can also be defined as the aberration of the optical path determined from finite ray trace data \cite{8}. The optical path differences (OPDs) between individual rays passing through the aperture and the chief ray are determined at the exit pupil of the system. In that case, surface contributions can be defined according to OPDs evaluated at the arbitrary selected reference spheres. Every choice of reference spheres, that ensures the additivity and provides valuable insights about the optical system, is feasible.  In optical design software a single set of rays, by default referred to normalized pupil coordinates at the entrance pupil \cite{9}, is traced through the system. Hence, the shape of the beam footprint and the intersection coordinates are distorted as rays propagate to intermediate reference spheres. In this case extrapolation and interpolation routines have to be used to find surface contributions to the total aberration assigned to unique aperture shape and intersection coordinates at the local reference spheres of each surface.\\ 
	In the following, we propose an alternative method to find surface-by-surface wave aberration contributions, utilizing the trace of multiple ray sets. Separate sets of rays are aimed at the unique coordinates on the respective reference spheres  and the change of the wavefront is evaluated. The optical system is divided into segments of the optical path measured along the chief ray with intermediate reference spheres located at each surface. Hence, surface contributions represent the wavefront change after each segment and are further divided with respect to their phenomenological origin into intrinsic, induced and transfer components. The intrinsic part results from refraction of an ideal wavefront at the surface. It is therefore independent of the rest of the system. Induced and transfer parts are both the effect of incoming aberrations and transverse deviations of the ray intersection points measured at the reference spheres. Hence, they both describe dependencies between individual surfaces. The method is based on the procedure used by Hoffman for validation of analytical results \cite{10}.\\ 
	Certain adjustments to the concepts known from the classical analytical approach have been made. Hence, in section \ref{sec:2} selected definitions of references are explained. Section \ref{sec:3} describes the method for determining surface contributions and respective components from the ray trace data of multiple ray sets.  Here, only systems with near circular apertures are considered for two reasons. Firstly, in section \ref{sec:4} the comparison with the results of the aberration function expansion assigned to the circular pupil, are presented. Secondly, obtained error maps are decomposed into Zernike Fringe polynomial set defined on the unit circle. Hence, for the first time additive Zernike surface contributions are obtained. This is discussed in section \ref{sec:5}.  Finally in section \ref{sec:6} an example of a single mirror system with freeform corrector plate is shown and the applicability of the method is demonstrated.\\
	
	\section{Notation}
	\label{sec:2}
	The aberration function W($\vec{H}, \vec{\rho}$) is commonly defined as the deviation from the ideal wavefront in the exit pupil, with the field vector $\vec{H}$ and the pupil vector $\vec{\rho}$ as arguments. The field vector is located at the object plane and defines the point source from which the wave originates. The pupil vector defines the coordinates of the point in which a particular ray intersects the exit pupil. Hence, each ray is described by its normalized coordinates in the field and the pupil.\\   
	In the numerical approach presented here, field points are considered separately. Rays originating from individual field points are therefore defined by their transverse intersection coordinates with the pupil ($\rho_x\,\rho_y$). The exit reference sphere is constructed upon the chief ray and the optical path for each ray is determined. The optical path of the chief ray is taken as the reference and the difference is calculated\\
	\begin{equation}\label{eq:1}
		OPD = OPL_{ref} - OPL_{ray}.
	\end{equation}
	The wavefront error is defined as a map of OPDs for all traced rays, scaled by the wavelength
	\begin{equation}\label{eq:2}
		W_{tot}(\rho_x,\rho_y) = \frac{OPD(\rho_x,\rho_y)}{\lambda}.
	\end{equation}
	\\
	\subsection{Pupil vector}
\begin{figure}[h]
		\begin{center}
			\begin{tabular}{c}
				\includegraphics[width=8cm]{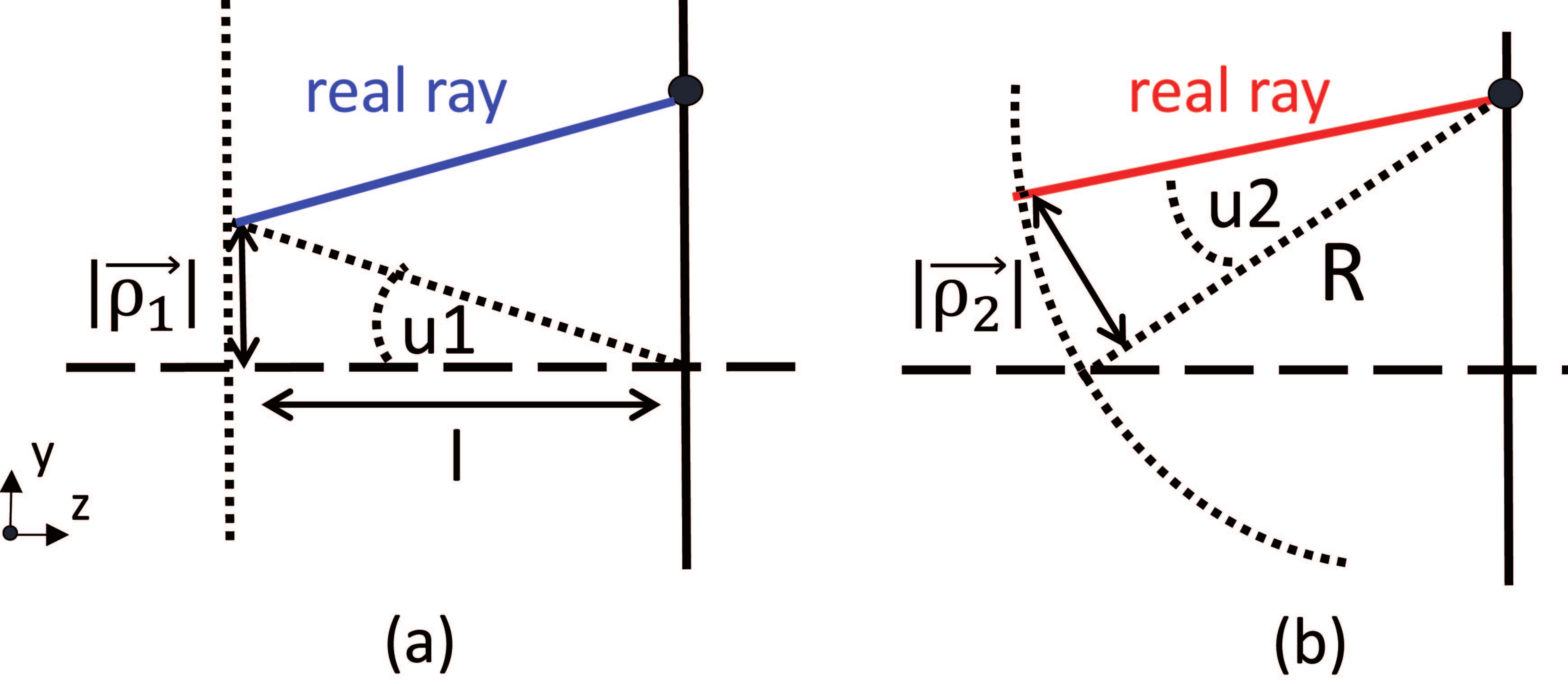}
			\end{tabular}
		\end{center}
		\caption 
		{ \label{fig:1}
			Locating the coordinate system of pupil vector (a) on the pupil plane and (b) on the pupil sphere oriented perpendicularly to the chief ray.} 
	\end{figure}

	In the aberration theory of axially symmetric systems the pupil vector is typically located on the equidistant grid in the exit pupil plane \cite{1}. The ideal path of each ray is then defined in the common set of coordinates in the field and the pupil.  This interpretation of first-order optics is known as Gaussian optics or tangent ideal \cite{11}. Alternatively an equidistant grid of transverse coordinates  on the exit pupil reference sphere, termed canonical coordinates \cite{12}, can be selected. This choice is appropriate for the measurement of the departure from isoplanatism and is suitable for calculations based on Fourier optics. Comparison of both possibilities for locating pupil vector is shown in Fig. \ref{fig:1}. In the first case transverse pupil coordinates $(\rho_x ,\rho_y )$ are related to the paraxial ray angle $u$ by
	\begin{equation}\label{eq:3}
		\frac{\vec{|\rho_1|}}{l} = \tan u_1,
	\end{equation}
	where $l$ is the distance between pupil and image planes and $\vec{|\rho_1}|$ is the length of the pupil vector measured in the transverse intersection coordinates.\\
	In the latter case point of the same transverse coordinates is placed at the pupil sphere oriented perpendicularly to the chief ray, therefore
	\begin{equation}\label{eq:4}
		\frac{\vec{|\rho_2|}}{R} = \sin u_2,
	\end{equation}
	where $R$ is the radius of the reference sphere defined along the chief ray  and $\vec{|\rho_2}|$ is the length of the pupil vector measured in the transverse intersection coordinates.\\
	\subsection{Exit pupil sphere }
\begin{figure}[h]
		\begin{center}
			\begin{tabular}{c}
				\includegraphics[width=5cm]{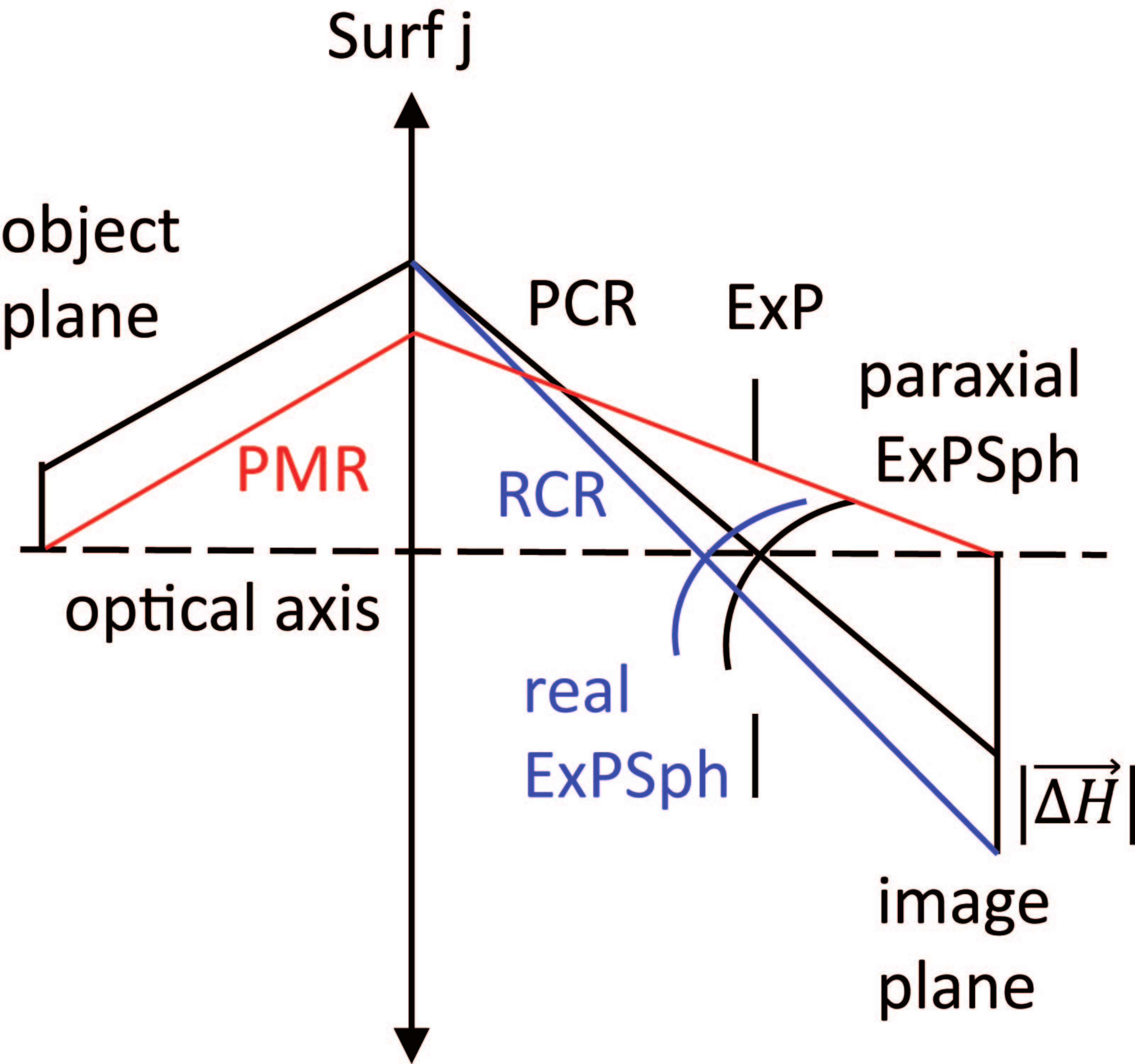}
			\end{tabular}
		\end{center}
		\caption 
		{ \label{fig:2}
			Difference between exit pupil spheres (ExPSph) referred to the paraxial chief ray (PCR) and the real chief ray (RCR). In the latter case the information about the chief ray aberration $|\Delta H|$ is not present in the aberrated wavefront. The diameter of the paraxial exit pupil (ExP) is determined by the paraxial marginal ray (PMR).} 
	\end{figure}

\begin{figure}[h]
		\begin{center}
			\begin{tabular}{c}
				\includegraphics[width=6cm]{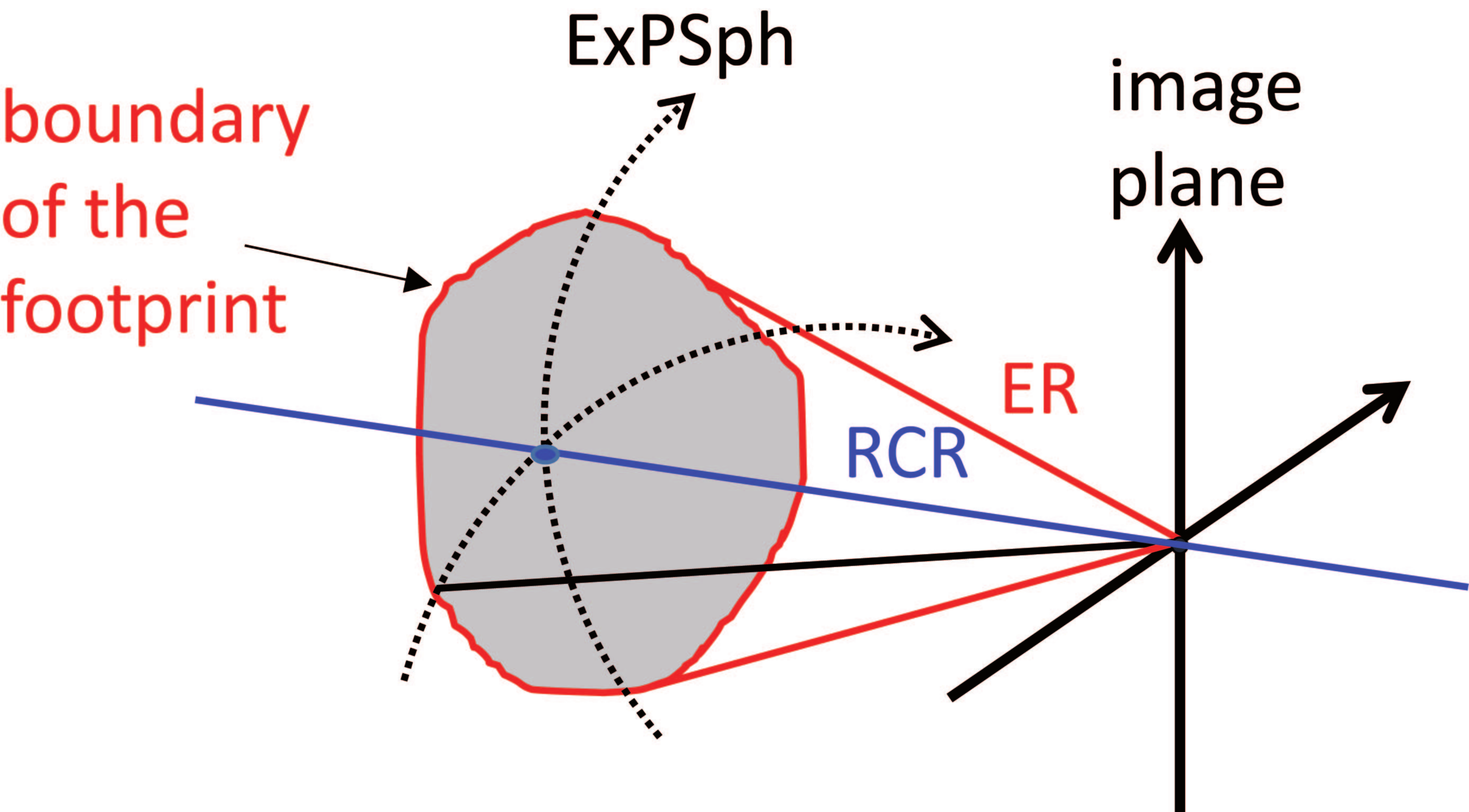}
			\end{tabular}
		\end{center}
		\caption 
		{ \label{fig:3}
			Preliminary trace of edge rays (ER) to determine the normalization aperture radius of the exit pupil sphere (ExPSph) located upon the real chief ray (RCR).} 
	\end{figure}

	For the axially symmetric systems the location and the diameter of the exit pupil sphere can be determined by tracing the paraxial chief ray and the marginal ray. The numerical results describe then the deviation from the ideal imaging. In the case of systems with broken symmetry, the exit pupil sphere is centered upon the intersection of the real chief ray with the image plane and located in the point of intersection with the reference optical axis. Hence, it is determined independently for each field point and the information about aberrations of the chief ray is lost. Figure \ref{fig:2} illustrates the difference in location of both exit pupil spheres in case of axially symmetric system. \\
	 \\
	To determine the normalized pupil coordinates the definition of the normalization radius of the exit pupil sphere is necessary. In the system with no symmetry constraints the paraxial marginal ray is not available for that purpose. Moreover, if the set of rays with the initially circular footprint shape is traced through the system, the footprint shape is in general no longer circular in the exit pupil. Consequently, a method to unambiguously define the normalization radius $R_{norm}$ for the beam with near-circular boundary shape is required. This is done with the preliminary trace of rays sampled around the edge of the beam [Fig. \ref{fig:3}]. Afterwards, the intersection points with the exit pupil sphere are calculated and the root-mean-square distances to the chief ray intersection point is evaluated
	\begin{equation}\label{eq:5}
		R_{norm}= \frac{1}{N}\sqrt{\sum_{i=1}^{N}(x_{ER}^{i} - x_{RCR})^2 + (y_{ER}^{i} - y_{RCR})^2}.
	\end{equation}
	The same method is used for radiometry calculations of optical systems with non-circular aperture stops \cite{13}.\\
	
	\subsection{Intermediate image planes}
		\begin{figure}[h]
		\begin{center}
			\begin{tabular}{c}
				\includegraphics[width=6cm]{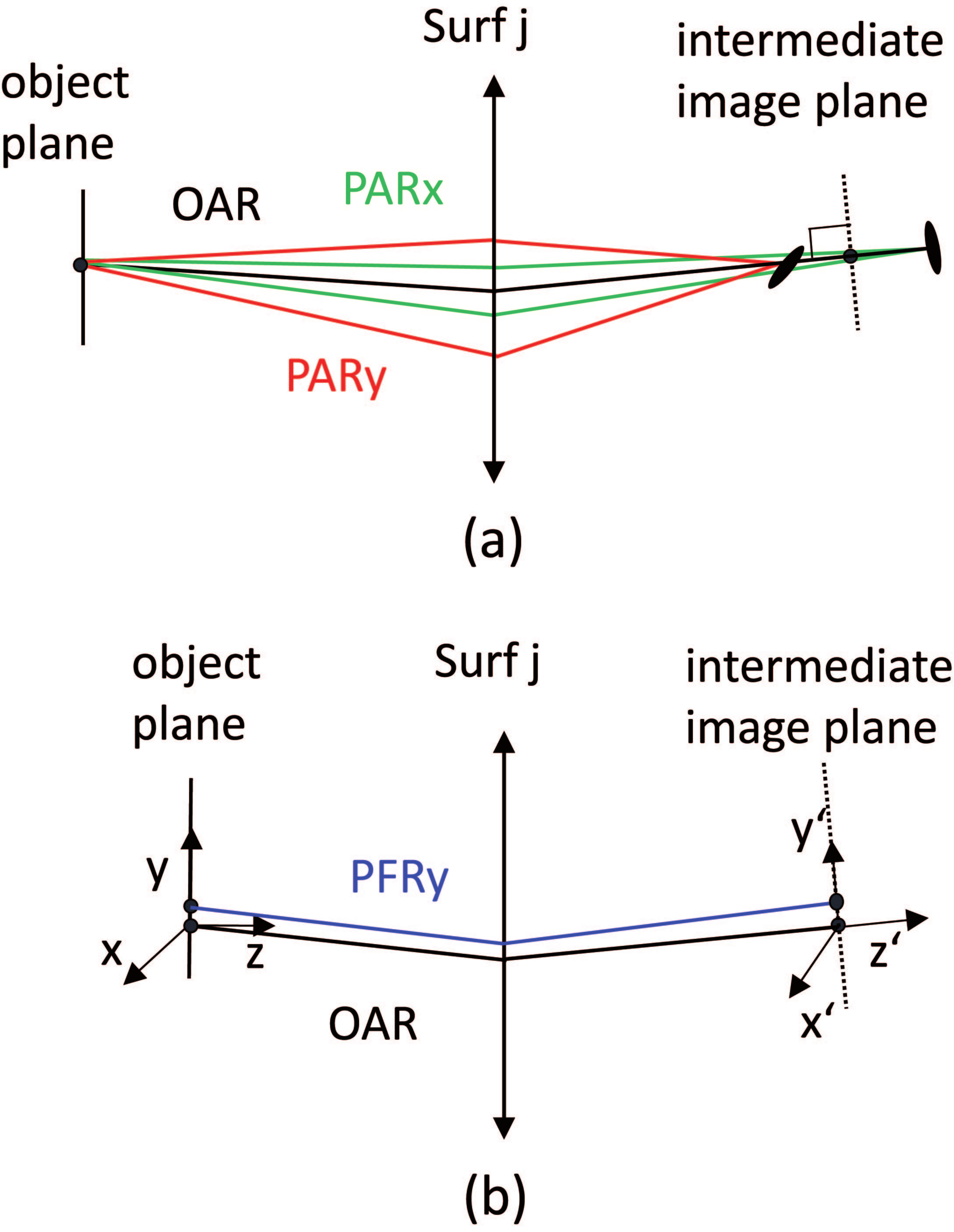}
			\end{tabular}
		\end{center}
		\caption 
		{ \label{fig:4}
			Construction of intermediate image planes with trace of parabasal rays. (a) Location and orientation determined with trace of two arbitrary, orthogonally oriented pairs of rays PARy and PARx.(b) Orientation of local transverse axes determined with  trace of parabasal ray in field PFRy launched from the tangential axis of the object plane.} 
	\end{figure}
	If surface contributions are to be determined, the definition of intermediate image planes is necessary. Individual image points are then defined as the points of intersection of real chief rays and intermediate image planes. In the implemented method centers of intermediate image planes are located by the base ray \cite{14} specified here as the optical axis ray. The optical axis ray (OAR), similarly as in nodal aberration theory \cite{2}, corresponds to the ray that connects the center of the field with the center of the aperture stop. The shift of aberration field quantified in nodal aberration theory by the vector $\vec{\sigma}$ is not considered. The exact location of the image planes is found from the trace of parabasal aperture rays [Fig. \ref{fig:4}(a)]. The fraction of the normalized pupil coordinates is selected so that the linear dependency in the aperture is fulfilled.  Since emerging wavefronts are in the general case astigmatic, two arbitrary, orthogonally oriented pairs of rays are traced. The intersection points of perpendicular pairs of rays are projected onto the optical axis ray and the middle point is found. This corresponds to the location of the circle of least confusion and is a suitable choice to balance intermediate astigmatism, which is especially pronounced in case of titled mirror systems. The orientation of intermediate image planes is chosen perpendicular to the optical axis ray. It is important to note that the object plane as well as the final image plane are not necessarily oriented perpendicular to the optical axis ray for non-telecentric imaging systems. \\
	Similarly, a parabasal ray in the field is traced to find the orientation of the local transverse axes of the image planes [Fig. \ref{fig:4}(b)]. This also allows for finding the reference location of individual image points. Hence, the distortion is defined separately from the wave aberration, as the transverse error on every intermediate image plane \cite{15}
	\begin{equation}\label{eq:6}
		Distortion = 100 \times\frac{\vec{|r_{chief}|}-\vec{|r_{ref}|}}{\vec{|r_{ref}|}}.
	\end{equation} 
	The definition of intermediate images, based on trace of rays near to the optical axis ray, assures the convergence to the paraxial optics in case of axially symmetric systems. \\

	\section{Definition of surface contributions}
	\label{sec:3}
	\subsection{An alternative division of the total wave aberration}

\begin{figure}[h]
		\begin{center}
			\begin{tabular}{c}
				\includegraphics[width=7cm]{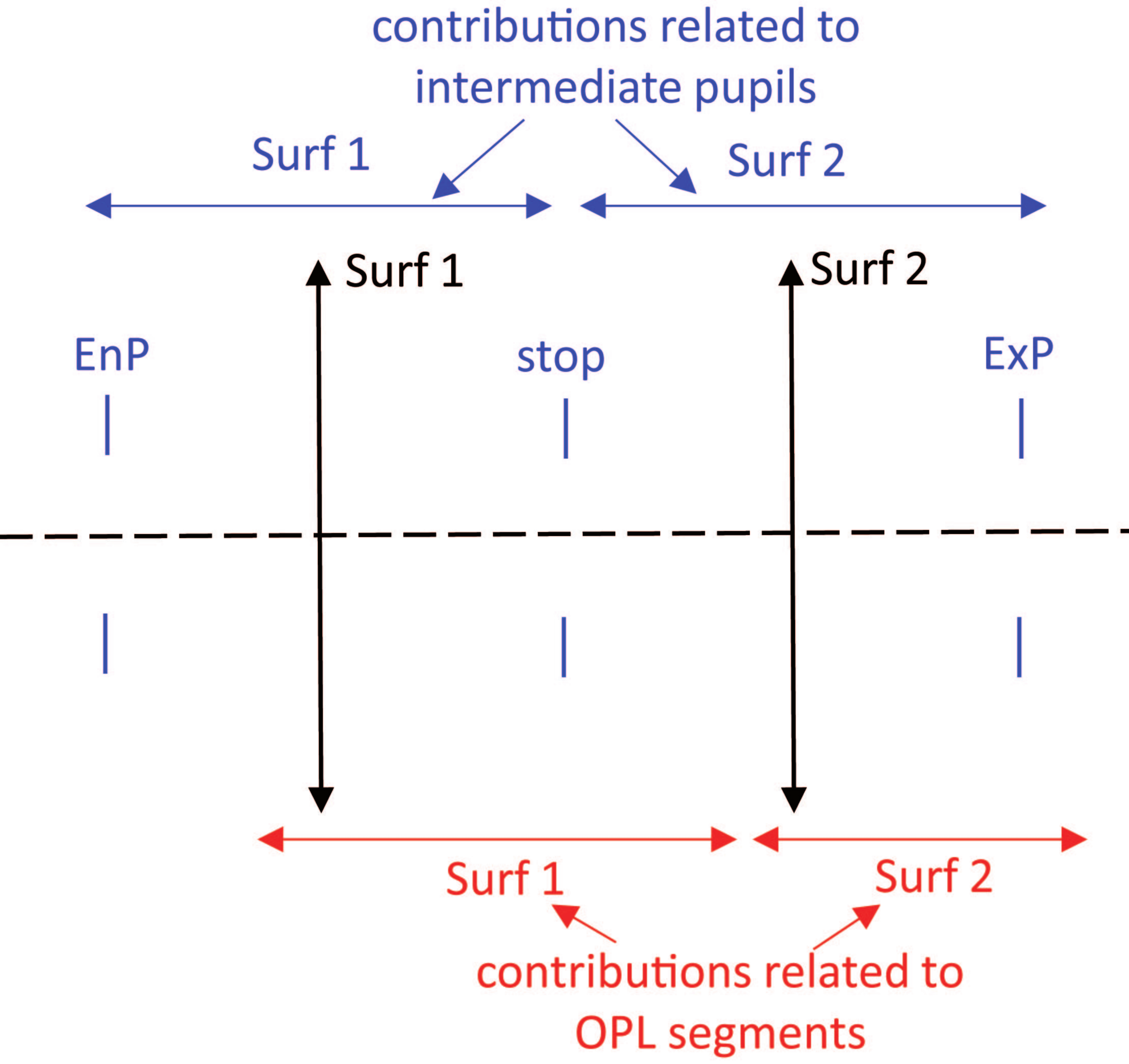}
			\end{tabular}
		\end{center}
		\caption 
		{ \label{fig:5}
			Alternative definition of surface contributions to the total wave aberration. Instead of dividing the system into subsystems bounded by intermediate pupils (EnP, stop, ExP) (blue color), surface contributions are defined as the segments of the optical path measured along the chief ray from the surface of interest until the subsequent surface (red color).} 
	\end{figure}

	If real chief rays are considered the construction of the pupil with unique location and size for all field points is not possible. Hence, in the presented method the concept of intermediate pupils is skipped and an alternative division of the total wave aberration is applied [Fig. \ref{fig:5}]. 
	Instead of dividing the system into subsystems with individual entrance and exit pupils, the separation into segments of the optical path measured along the chief ray from the surface of interest until the subsequent surface, is applied. Surface contributions are bounded by the entrance reference spheres located directly at the intersection of the real chief rays with subsequent surfaces. In case the surface is near the focus, it is ambiguous to set reference sphere near strong caustic region. Nevertheless, this only concerns intermediate results. The total wave aberration is evaluated at the exit pupil reference sphere of the system as described in section \ref{sec:2}.1. The definition of surface contributions according to construction parameters instead of physical properties gives the lens designer additional insights into limitations and possibilities of the optical system.
	
	\subsection{Componets of surface contributions determined from the trace of multiple ray sets}
\begin{figure}[h]
		\begin{center}
			\begin{tabular}{c}
				\includegraphics[width=\linewidth]{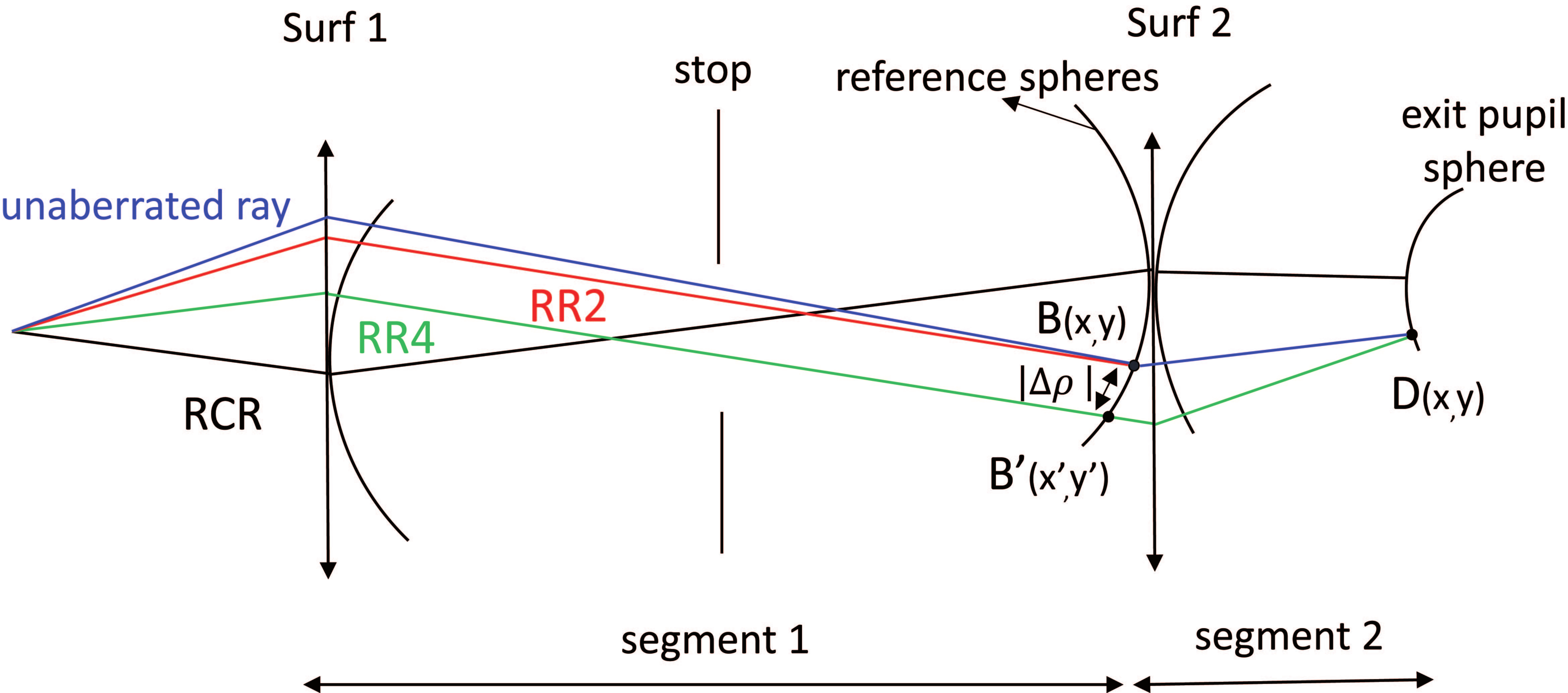}
			\end{tabular}
		\end{center}
		\caption 
		{ \label{fig:6}
			Complete surface contributions to the total OPD determined from the trace of mutiple rays (RR2 and RR4).} 
	\end{figure}

\begin{figure}[h]
		\begin{center}
			\begin{tabular}{c}
				\includegraphics[width=3.8cm]{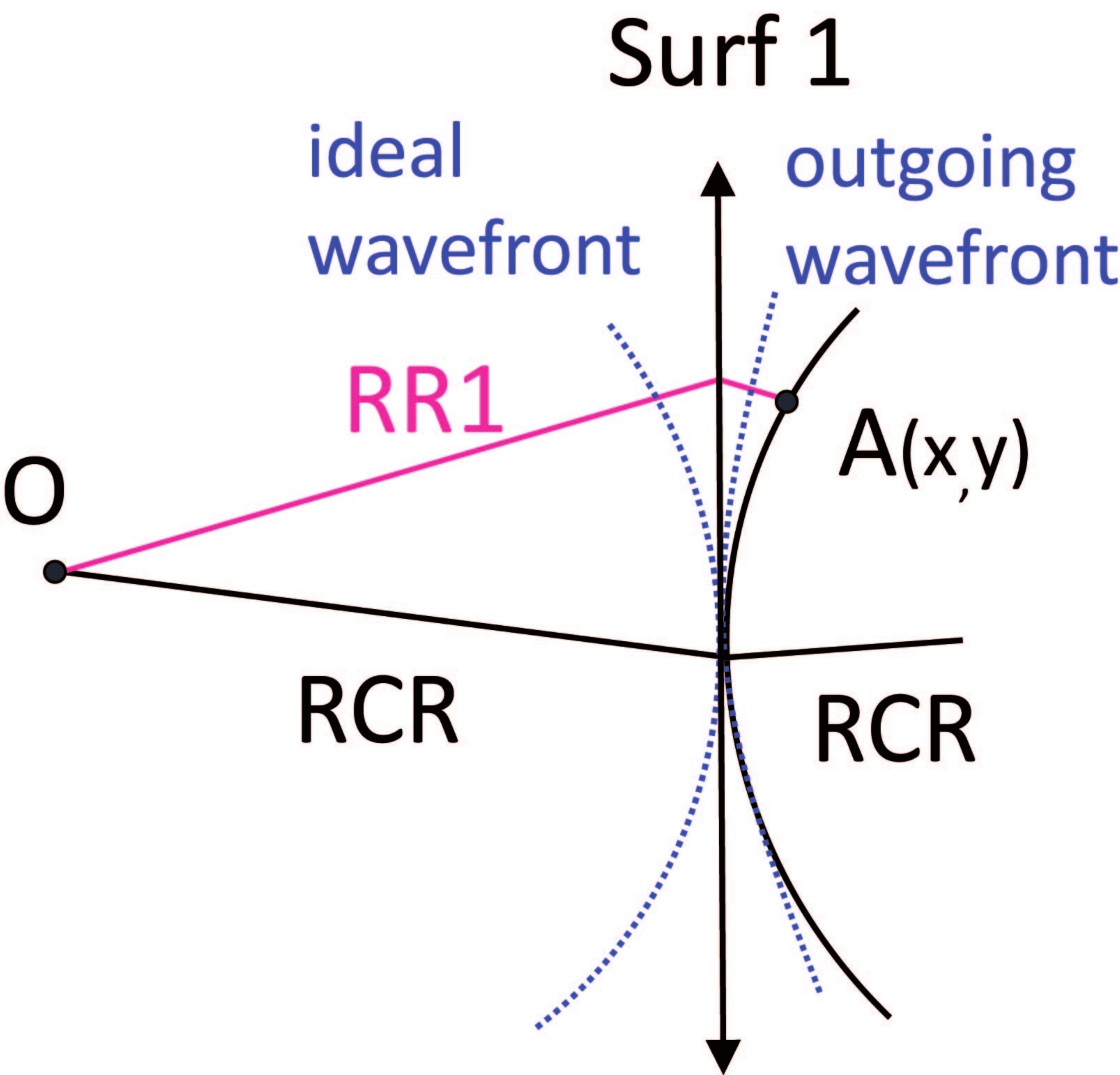}
			\end{tabular}
		\end{center}
		\caption 
		{ \label{fig:7}
			Ray RR1 is traced to determine the component of the total OPD resulting from the refraction on the first surface} 
	\end{figure}

\begin{figure}[h]
		\begin{center}
			\begin{tabular}{c}
				\includegraphics[width=7cm]{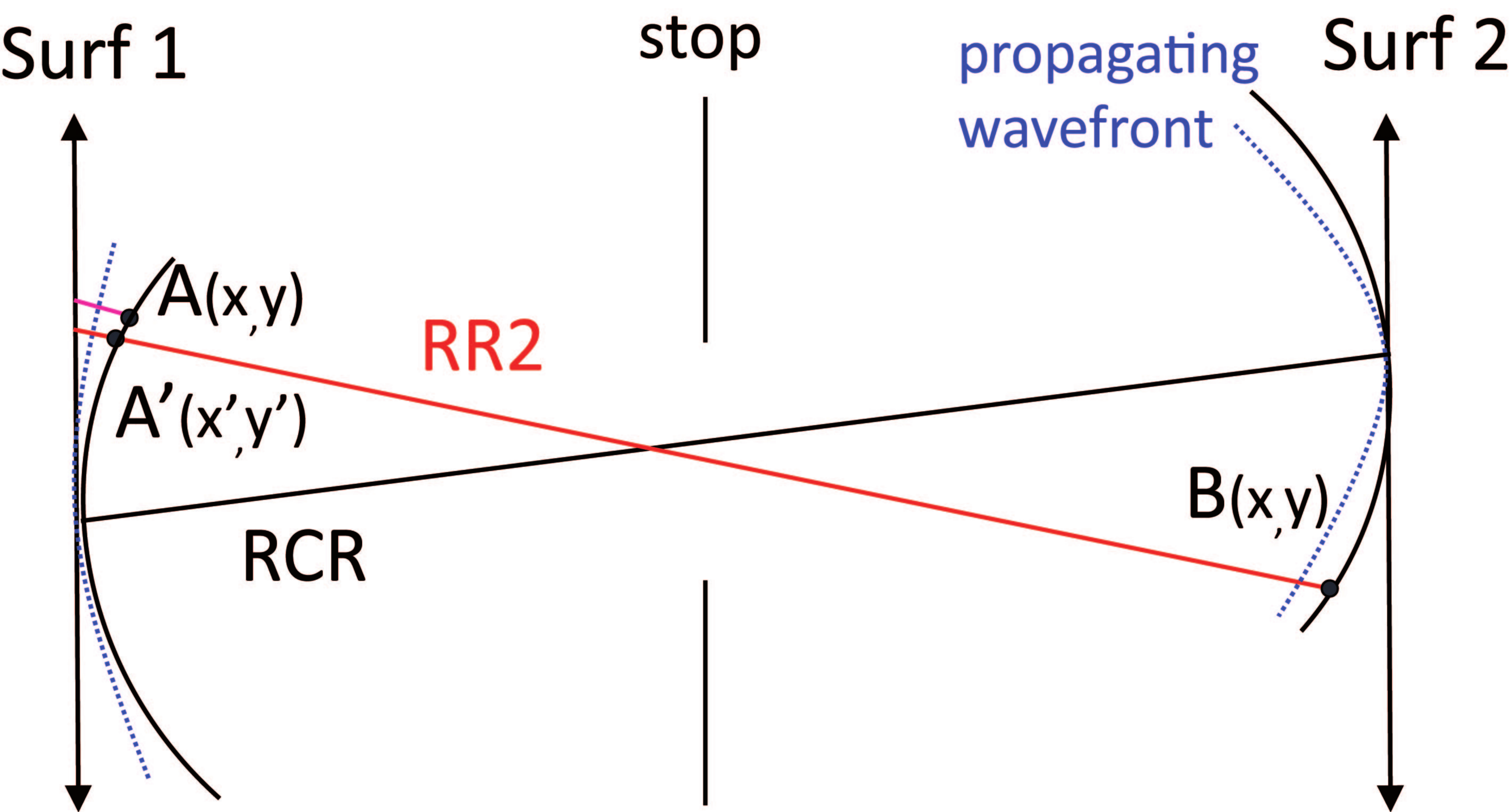}
			\end{tabular}
		\end{center}
		\caption 
		{ \label{fig:8}
			Ray RR2 is traced to determine the transfer component of the total OPD resulting from propagation between surfaces.} 
	\end{figure}

	The total wavefront error is defined as the map of OPDs measured along real rays traced to the exit pupil sphere of the system. Transverse pupil coordinates $(\rho_x ,\rho_y )$  are assigned to the equidistant cartesian grid $(x,y)$ on the final reference sphere. If only one set of rays is traced, the grid created on an arbitrary intermediate reference sphere is distorted, which we indicate with primed coordinates $(x’,y’)$. This is solved by tracing multiple sets of rays aimed at the equidistant grids $(x,y)$ of local coordinates on each reference sphere.  Consequently, instead of measuring the OPD along a single ray up to the exit pupil sphere, multiple rays are used. This can be thought of as evaluating wavefront errors after each surface at the similar set of coordinates on the reference spheres. Hence, wavefront errors can be subtracted from each other to find the change caused by a particular segment of the system. This is explained for a single OPD value in the Fig. \ref{fig:6}.\\ 
	The contribution of the first surface is defined as the OPD at point $B$. The contribution of the second surface is found from the difference between the OPD on points $D$ and $B$ on similar transverse coordinates on the entrance and the exit reference sphere of the segment
	\begin{equation}\label{eq:7}
		OPD_{S2} = OPD_{D} - OPD_{B}.
	\end{equation} \\
	Additivity is therefore preserved, since the sum of both surface contributions is equivalent to the total OPD calculated at point $D$ along a single ray
	\begin{equation}\label{eq:8}
		OPD_{total} = OPD_{S1} + OPD_{S2} = OPD_{D}.
	\end{equation} 
	There is no real ray intersecting the points $B$ and $D$. One can imagine an extrapolated parabasal ray connecting uniform coordinates on all reference spheres. \\
	The advantage of utilizing multiple ray sets is that the transverse pupil aberration $-\vec{|\Delta\rho|}$, which negative sign is due to the difference with the analytical approach [section \ref{sec:4}], is incorporated in the contribution of the second surface
	\begin{equation}\label{eq:9}
		-\vec{|\Delta\rho|} = [(B'_x-B_x), (B'_y-B_y)].
	\end{equation} \\
	This allows to extract the induced effect defined here as a result of incoming aberrations and pupil distortion. Hence, the surface contribution is further divided into intrinsic and induced parts resulting from refraction on the surface and the transfer component, which is present due to the propagation of the aberrated wavefront in the free space.\\
	In case of the first segment of the system from Fig. \ref{fig:6}, the entering wavefront is ideal. Since there are no incoming aberrations, the pupil distortion is of no effect. The refraction on the surface is of purely intrinsic type. It is therefore enough to trace one ray to point $A$ at the equidistant grid $(x,y)$ of the exit reference sphere located on the first surface [Fig. \ref{fig:7}].
	\\
	\\
		To determine the transfer component the second ray is traced to the equidistant grid at the second surface entrance sphere [Fig. \ref{fig:8}].
	
\begin{figure}[h]
		\begin{center}
			\begin{tabular}{c}
				\includegraphics[width=3.7cm]{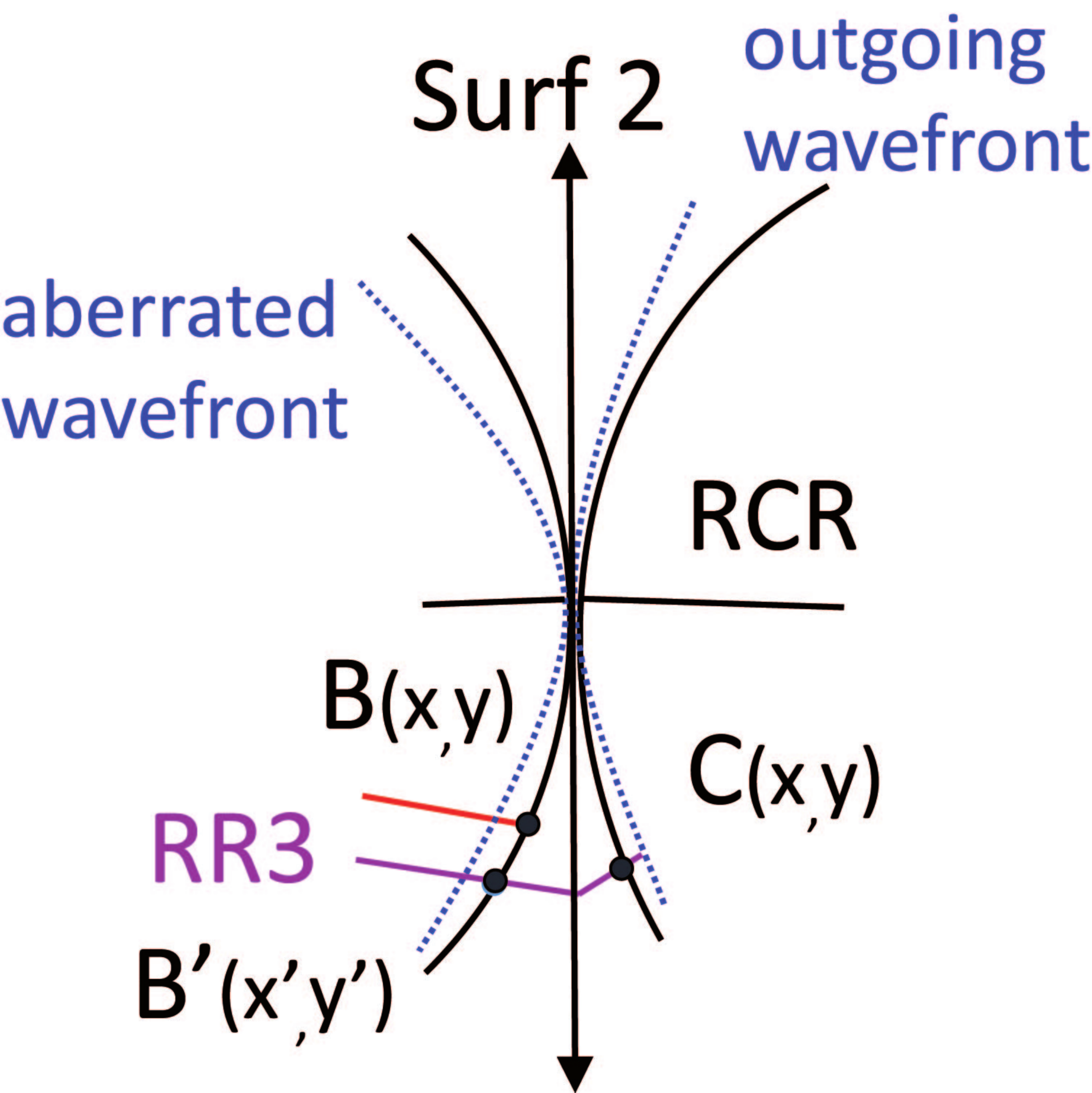}
			\end{tabular}
		\end{center}
		\caption 
		{ \label{fig:9}
			Ray RR3 is traced to determine the component of the total OPD resulting from refraction on the second surface.} 
	\end{figure}

\begin{figure}[h]
		\begin{center}
			\begin{tabular}{c}
				\includegraphics[width=4.2cm]{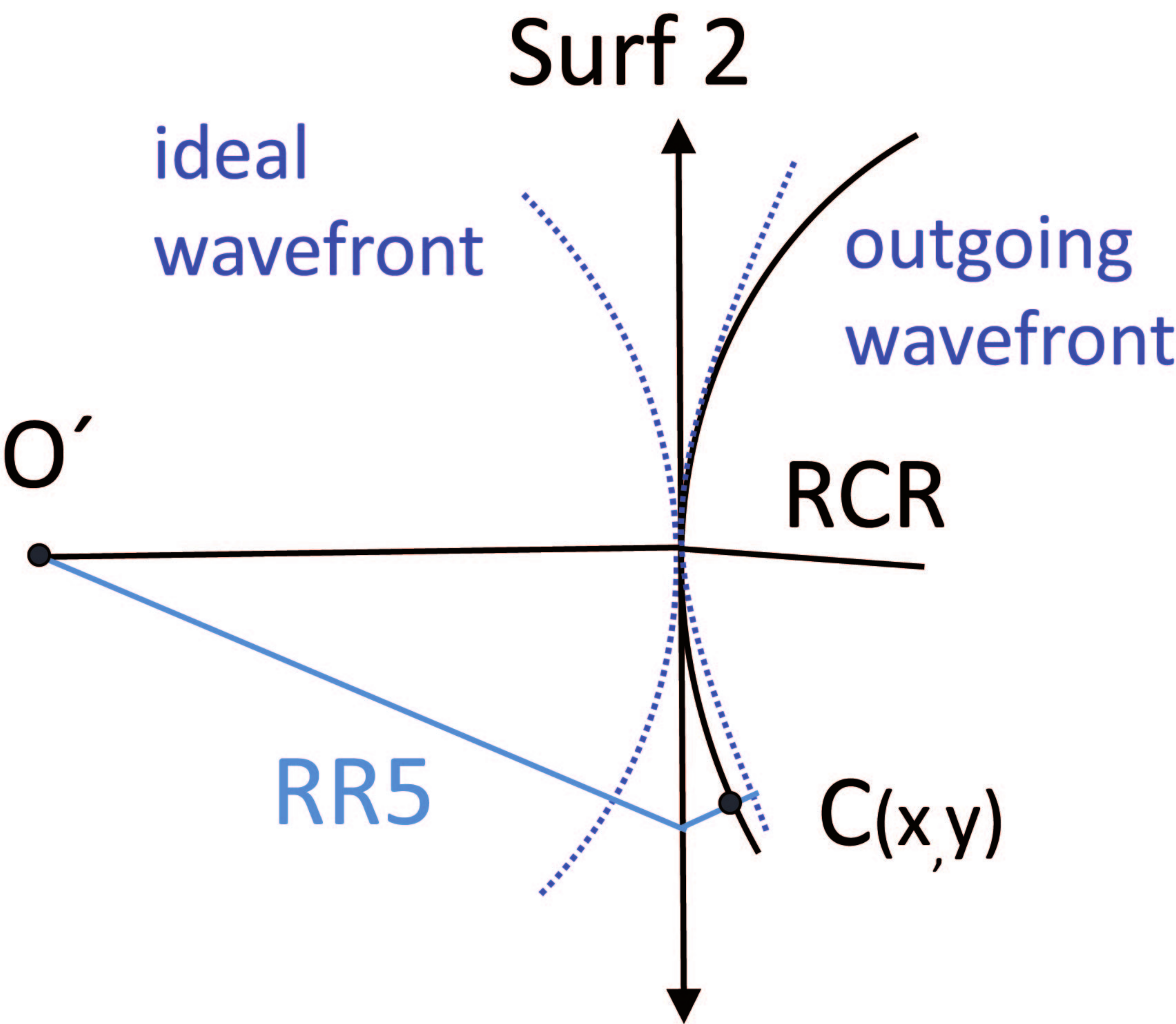}
			\end{tabular}
		\end{center}
		\caption 
		{ \label{fig:10}
			Ray RR5 is traced to determine the intrinsic component of the total OPD resulting from refraction of an ideal wavefront on the second surface.} 
	\end{figure}

\begin{figure}[h]
		\begin{center}
			\begin{tabular}{c}
				\includegraphics[width=3.5cm]{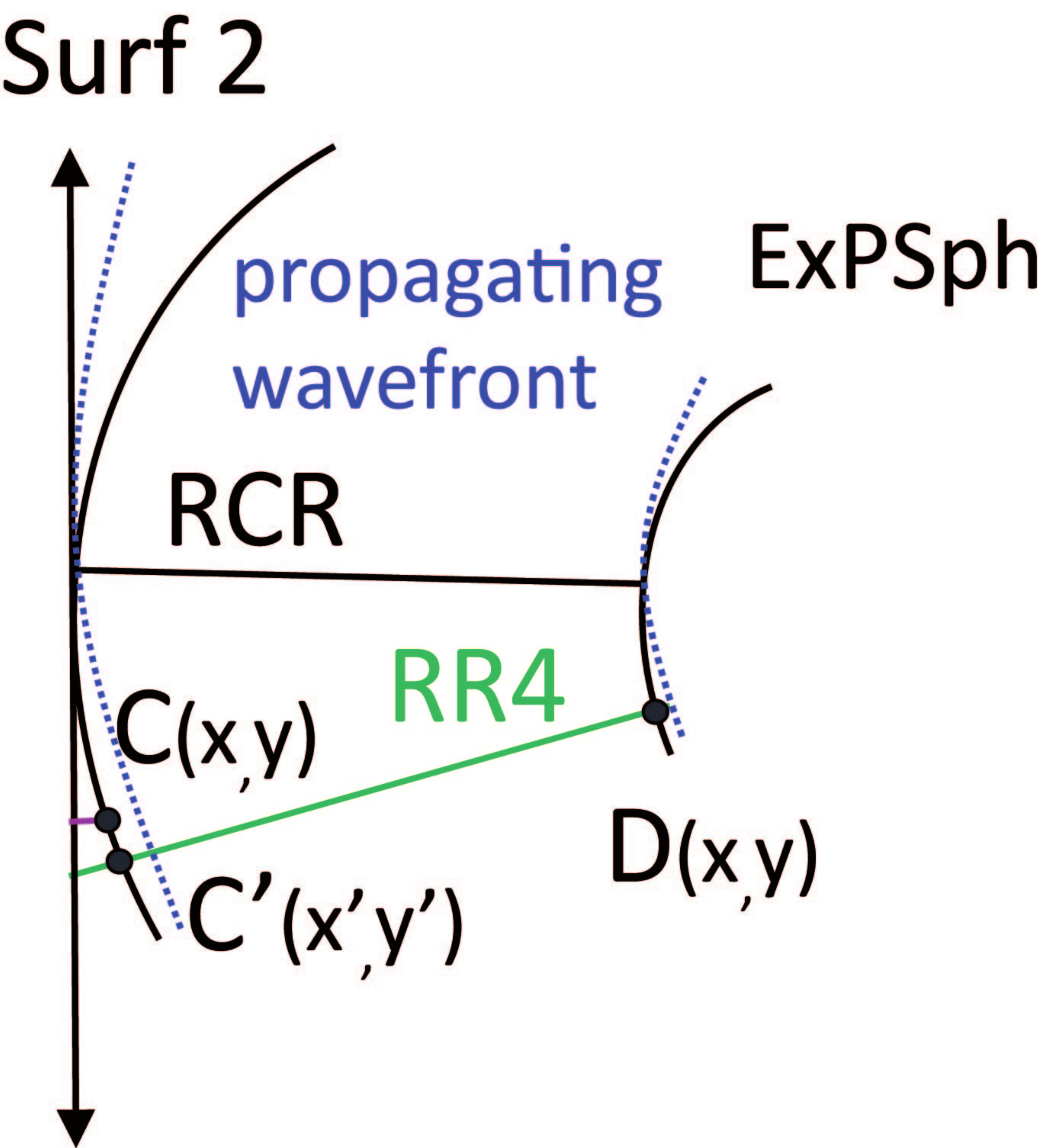}
			\end{tabular}
		\end{center}
		\caption 
		{ \label{fig:11}
			Ray RR4 is traced to determine the transfer component of the total OPD resulting from propagation to the exit pupil of the system.} 
	\end{figure}

	The transfer part $OPD_{T1}$ is then found by subtracting the OPD at point $A$ from the OPD at point $B$ 
	\begin{equation}\label{eq:10}
		OPD_{T1} = OPD_{B} - OPD_{A}.
	\end{equation} \\
	Defining the transfer component in this way takes into account the transverse pupil aberration measured at the reference sphere between the points $A’$ and $A$. Hence, the transfer component is a combined effect of pupil distortion caused by incoming aberrations of the first surface and the deformation arising upon propagation of the aberrated wavefront between surfaces. Since the propagation of an ideal wavefront in the free space only changes the scaling without introducing any aberration, the transfer component is considered as the part of induced effect. This is different than in the classical division according to individual pupils of the surfaces [Fig. \ref{fig:5}]. In that case, the transfer term is not distinguished and divided between intrinsic and extrinsic parts of successive surfaces \cite{10}. The advantage of separating the transfer component is that it directly refers to the design variables of the system.\\
	The OPD introduced by the first surface is then determined by summing up the transfer and the refraction parts.\\
	The complete effect of refraction of the incoming aberrated wavefront on the second surface $OPD_{R2}$ is found by subtracting the contribution of the first surface from the OPD at point $C$ at the exit reference sphere located on the second surface [Fig.\ref{fig:9}]\\
	\begin{equation}\label{eq:11}
		OPD_{R2} = OPD_{C} - OPD_{B}.
	\end{equation}
	\\
	In order to measure the effect of refraction on the second surface independently from the rest of the system, the intrinsic component is introduced. The intrinsic part is a deformation of an ideal wavefront after refraction at the surface. Consequently, to determine the intrinsic OPD, an additional ray has to be traced until point $C$ [Fig. \ref{fig:10}].\\  This is realized by ignoring the first surface and tracing rays directly from the intermediate image location $O’$, defined as in section \ref{sec:4}.3 .\\
	The induced part from the refraction on the second surface is found by subtracting the intrinsic component from the complete effect of the refraction. It therefore measures the wavefront change due to the effect of incoming aberrations and pupil distortion 
	\begin{equation}\label{eq:12}
		OPD_{R2}^{ind} = OPD_{R2} - OPD_{R2}^{int}.
	\end{equation}
	The transfer component to the exit pupil of the system is found in analogy to the transfer part of the first surface contribution, [Fig. \ref{fig:11}].\\
	
	Hence, in order to determine all the components of surface contributions of the example system, with two surfaces and a remote exit pupil, the trace of five ray sets is necessary. First four ray sets are traced through the complete system with the equidistant grids located at the respective reference spheres. Last ray set is traced from the intermediate image location through the second surface. \\

	\section{Comparison with the analytical results}
	\label{sec:4}
	As mentioned in the introduction, it is possible to analytically determine surface contributions of axially symmetric systems up to the sixth order \cite{1}. In the following section results obtained with the proposed method are compared with wavefront error maps generated from aberration coefficients, calculated with the macro of Sasian \cite{16}.  For that purpose an axially symmetric, two mirror system represented schematically in Fig. \ref{fig:8} was investigated. Since the stop is located on the second surface, the exit pupil of the first surface contribution is located directly before the second mirror. Hence, the two divisions of the total wave aberration from Fig. \ref{fig:5} are equivalent. For the sake of comparison, the exit pupil spheres are constructed upon the paraxial chief ray. Therefore aberrations due to displacement of the field vector $\vec{|\Delta H|}$  are present in evaluated wavefront errors. \\
	\begin{figure}[h]
		\begin{center}
			\begin{tabular}{c}
				\includegraphics[width=5.2cm]{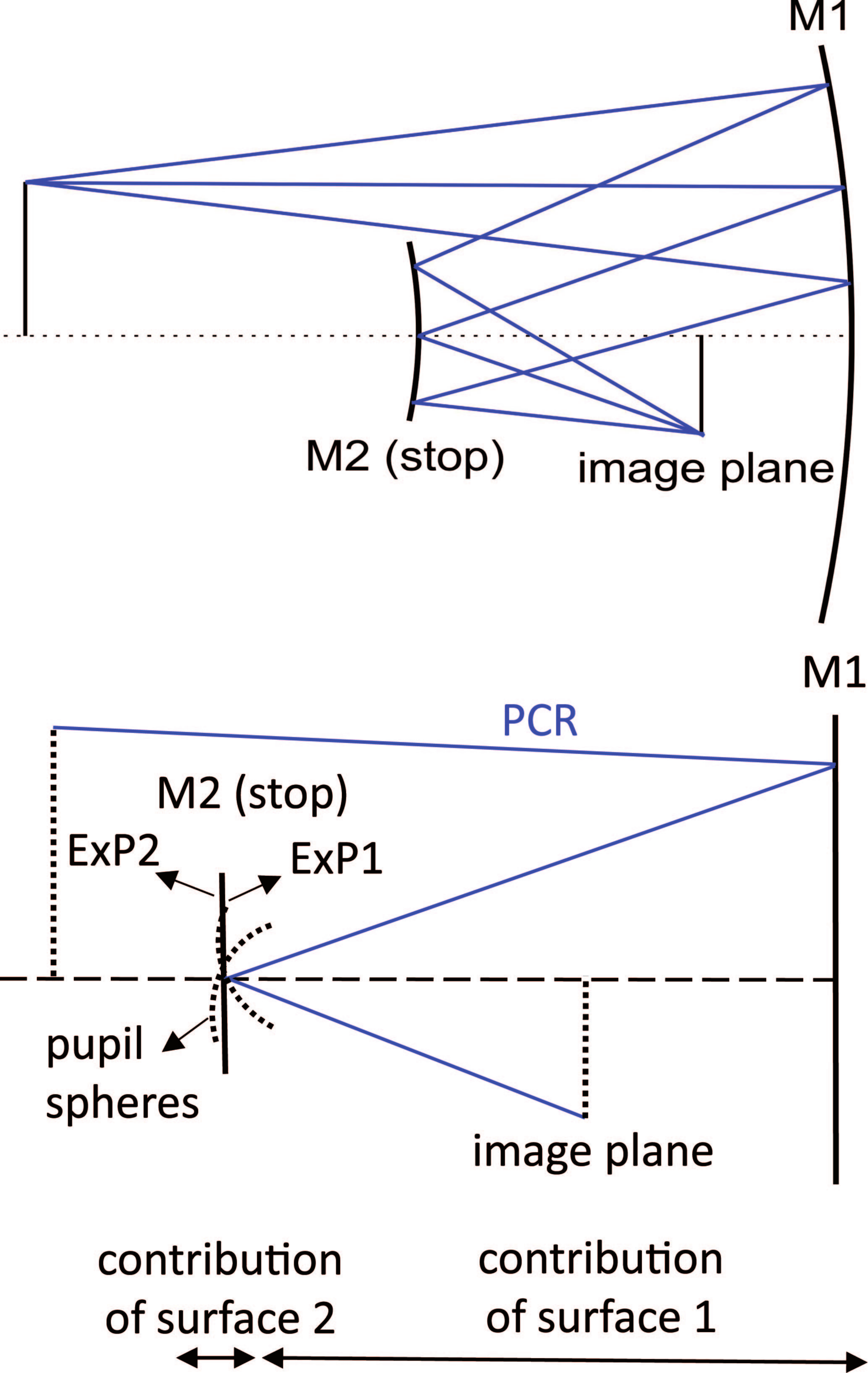}
			\end{tabular}
		\end{center}
		\caption 
		{ \label{fig:12}
			Two mirror system with the stop located at the second surface M2. Exit pupil of the first surface contribution ExP1 is located directly before the second mirror. The exit pupil of the second surface contribution is sequenced directly after second mirror ExP2. Hence, aberrations of the second surface are exclusively a result of the reflection from the second mirror. Since investigated system is axially symmetric, reference spheres are centered upon paraxial chief ray PCR.} 
			\end{figure}
	
	A comparison of results obtained with both methods shows that the distribution of surface contributions is changed [Figs. \ref{fig:9}(b) and (d)]. This is due to the difference in the location of the pupil vector. All wave aberration coefficients are evaluated for the pupil vector $\vec{\rho}$ located at the exit pupil plane of the system, whereas in the numerical approach, the OPDs are referred on the equidistant grids on the exit pupil spheres of respective surfaces, constructed upon the paraxial chief ray. It means that, in case of the first surface, the contribution is referred to the exit pupil of the first mirror which corresponds to the distorted vector $\vec{\rho}+\vec{\Delta\rho}$  at the pupil of the systems.  Hence, the redistribution of surface contributions is due to two factors. The use of pupil spheres instead of planes and the change in the location of the pupil vector from exit pupil to individual pupils of surfaces.\\
	The total aberrations of the system are affected only from the change in the pupil shape. The difference is therefore less significant as shown in Fig. \ref{fig:9}(f). \\
	\begin{figure}[h]
		\begin{center}
			\begin{tabular}{c}
				\includegraphics[width=\linewidth]{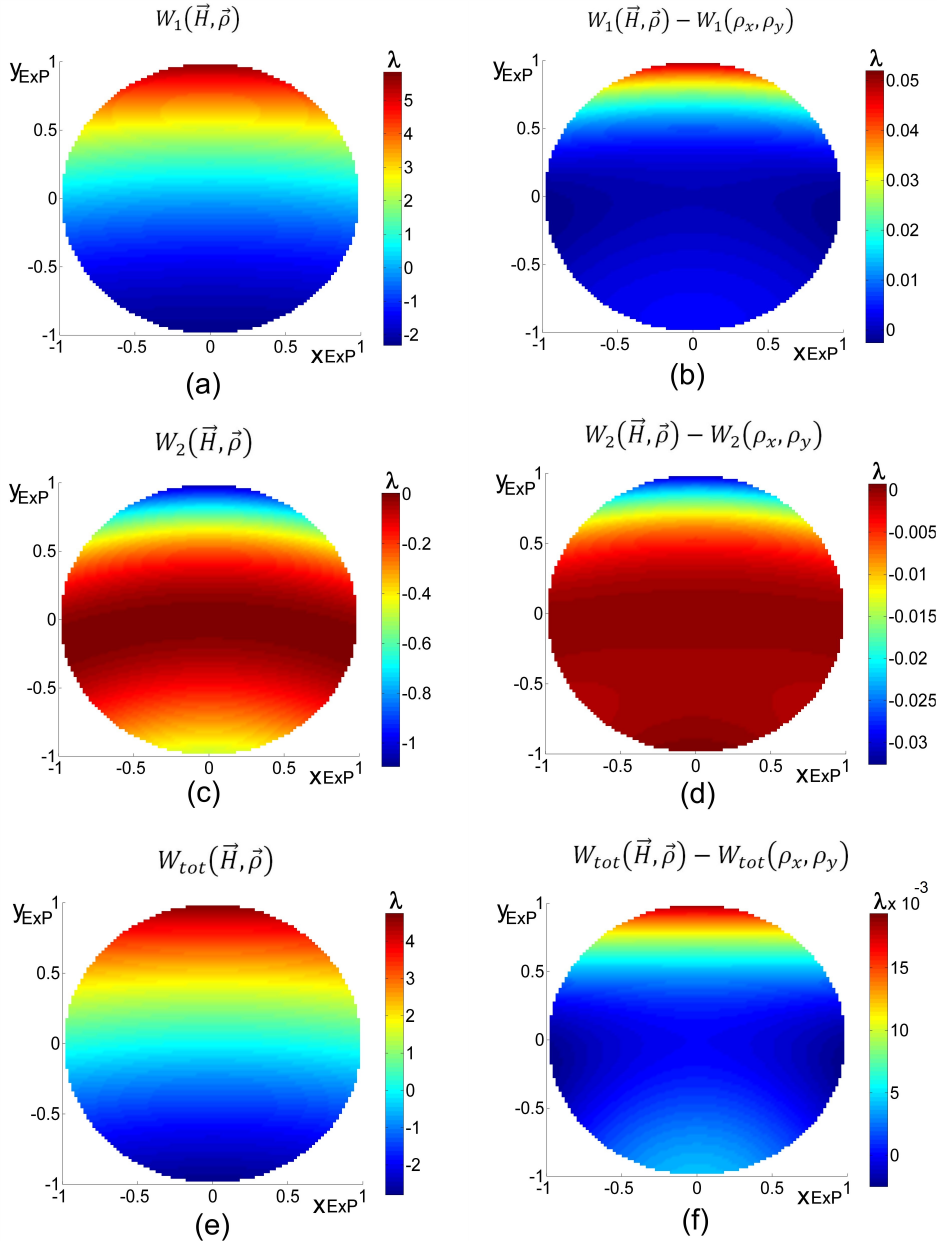}
			\end{tabular}
		\end{center}
		\caption 
		{ \label{fig:13}
			Comparison of wavefront error maps evaluated from wave aberration coefficients  and obtained with the proposed numerical method (a) contribution from first surface according to aberration coefficients, (b) the difference between analytical and numerical results, (c) contribution from second surface according to aberration coefficients, (d) the difference between analytical and numerical results, (e) the total wave aberration according to aberration coefficients, (f) residual difference between analytical and numerical total wavefront errors.} 
	\end{figure}
	The redistribution of surface contributions due to different position of equidistant pupil grid results in the sign difference of the induced component [Eq. (\ref{eq:9})] . In ref. \cite{1}, the induced part of the second surface up to sixth order depends on the fourth order incoming aberrations and third order transverse pupil aberration $\vec{\Delta\rho}^{(3)}$ of the first surface, measured with respect to the exit pupil of the system 
	\begin{equation}\label{eq:13}
		W_{2}^{(6, ind)}(\vec{H}, \vec{\rho}) =  W_{1}^{(4)}(\vec{H}, \vec{\rho}+\vec{\Delta\rho}^{(3)})-W_{1}^{(4)}(\vec{H}, \vec{\rho}).
	\end{equation}
	\\
	Analogously, in case of the example system for the numerical approach, the induced part of the second surface contribution with no restriction to the expansion order can be noted by
	\begin{equation}\label{eq:14}
		W_{2}^{(ind)}(\vec{H},\vec{\rho}) =  (W_{tot}(\vec{H},\vec{\rho}) - W_{1}(\vec{H},\vec{\rho}+\vec{\Delta\rho}))-W_{2}^{(int)}(\vec{H},\vec{\rho})
	\end{equation}.
	\\
	The term in the parenthesis is the complete contribution of the second surface defined as in the previous section [Eq. (\ref{eq:11})]. The transverse pupil aberration is included with a different subtraction order. The sign of induced components in Fig. \ref{fig:10} is changed as a result of the difference in sign of transverse pupil aberrations indicated in Eqs. (\ref{eq:13}) and (\ref{eq:14}).\\ Difference in shape is due to the transverse pupil aberration measured on the sphere see Fig. \ref{fig:10}(b) compared to a plane in Fig. \ref{fig:10}(a).
	\begin{figure}[h]
		\begin{center}
			\begin{tabular}{c}
				\includegraphics[width=\linewidth]{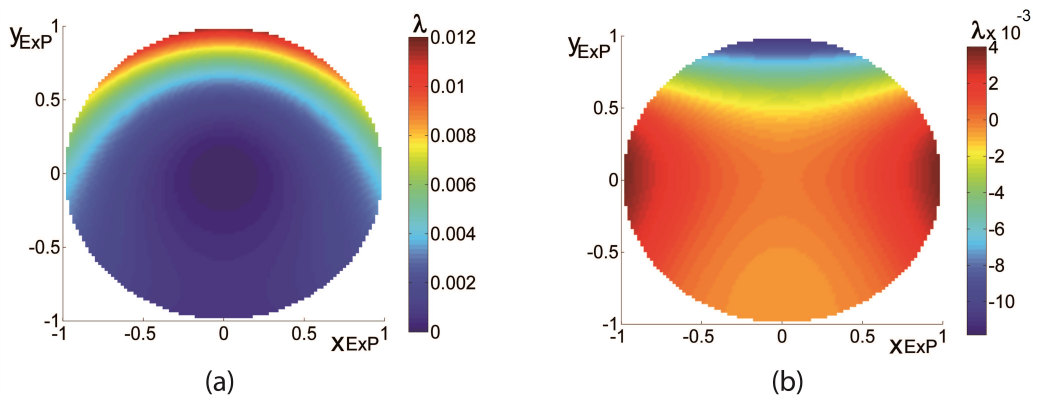}
			\end{tabular}
		\end{center}
		\caption 
		{ \label{fig:14}
			Comparison of second surface contribution induced components (a) extrinsic sixth order term, (b) induced component obtained numerically.} 
	\end{figure}
	\section{Wavefront errors decomposed into Zernike Fringe polynomials}
	\label{sec:5}
	The zernike fringe polynomials decomposition is the preferred method to analyze numerically acquired wavefront errors of systems with near circular apertures. Optical design programs typically offer analysis tools where Zernike coefficients are obtained from the data of a single ray set. This method refers the wavefront error to the single set of normalized coordinates, neglecting the aperture shape changes and the distortion of intersection coordinates with intermediate reference spheres. Hence, it is impossible to unambiguously determine Zernike coefficients assigned to each surface contribution from direct fit of the numerical results. The solution with the trace of multiple ray sets overcomes those inaccuracies. The wavefront error for each segment is obtained for the equidistant grid on the circular aperture. The normalization radius is defined for each reference sphere in a similar way as the normalization radius of the exit pupil [section \ref{sec:2}.2]. Moreover, as stated in Eq. (\ref{eq:8}) surface contributions are additive to the total wave aberration. Hence, the least square fit to the Zernike fringe polynomials is directly applicable
	\begin{equation}\label{eq:15}
		W_{tot}(\rho_x ,\rho_y )= \sum_{s}\sum_{j}c_{js}Z_j(\rho_x ,\rho_y ),
	\end{equation}
	where $j$ represents the number of coefficients used in the decomposition routine and $s$ is the index of surfaces. 
	Relations to Zernike terms obtained directly from the coefficients of the aberration function expansion are given in \cite{4}. For the method described here those relations are not directly transferable. Nevertheless, they might be helpful in the interpretation of evaluated Zernike coefficients. Hence, similar formulation to the one used by Fuerschbach et al. \cite{3}, relating Zernike terms by their corresponding aberration expansion order, is used. The difference is that Zernike aberrations are not adjusted by incorporating lower order terms. Up to the sixth order, five pairs of non-symmetric Zernike terms are obtained, namely, Zernike astigmatism (Z5/6), Zernike coma (Z7/8), Zernike elliptical coma or trefoil (Z10/11), Zernike oblique spherical aberration or secondary astigmatism (Z12/13), and Zernike sixth order aperture coma or secondary coma (Z14/15). Furthermore, rotationally symmetric terms of defocus (Z4), Zernike spherical aberration (Z9) and sixth order spherical aberration (Z16) are distinguished. This formulation allows to generate full field displays for selected pairs of Zernike terms, also in the reference to intermediate image planes. Hence the performance for different field points can be analyzed simultaneously.
	
	\section{Example}
	\label{sec:6}
	As an example a single spherical mirror system with a freeform corrector plate is investigated, illustrated in Fig. \ref{fig:11}(a). It is similar to the known Schmidt telescope with the difference that the stop is placed at the mirror. Each field point [Fig. \ref{fig:11}(b)] is therefore influenced differently by the corrector plate. The freeform element sag is described with Zernike Fringe coefficients 5 and 9.
	
	\begin{figure}[h]
		\begin{center}
			\begin{tabular}{c}
				\includegraphics[width=\linewidth]{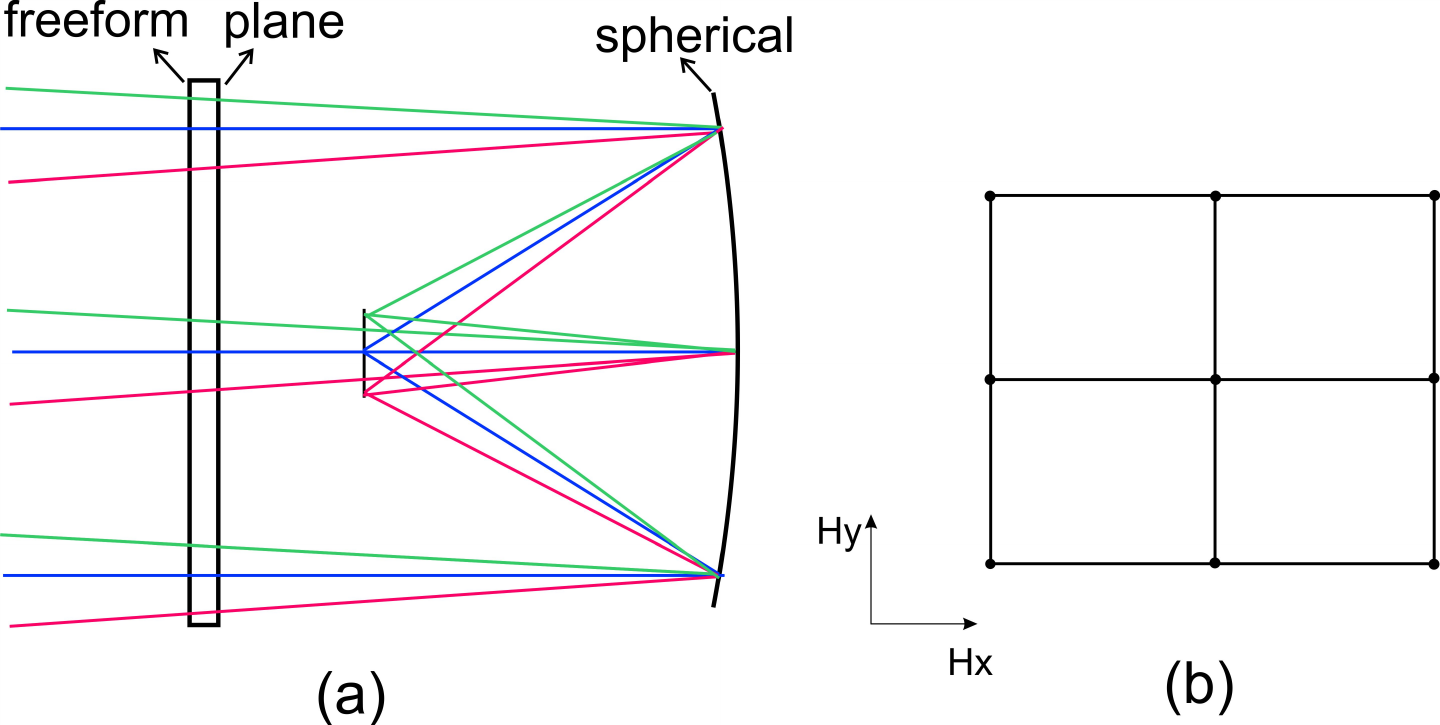}
			\end{tabular}
		\end{center}
		\caption 
		{ \label{fig:15}
			(a) Example, double plane symmetric system of single mirror with a freeform corrector plate, (b) square grid of field points in normalized field coordinates.} 
	\end{figure}
	The system is optimized for 9 field points with the minimum spot RMS criterion. Fig. \ref{fig:12} shows the obtained, inhomogeneous spots distribution.
	\begin{figure}[h]
		\begin{center}
			\begin{tabular}{c}
				\includegraphics[width=\linewidth]{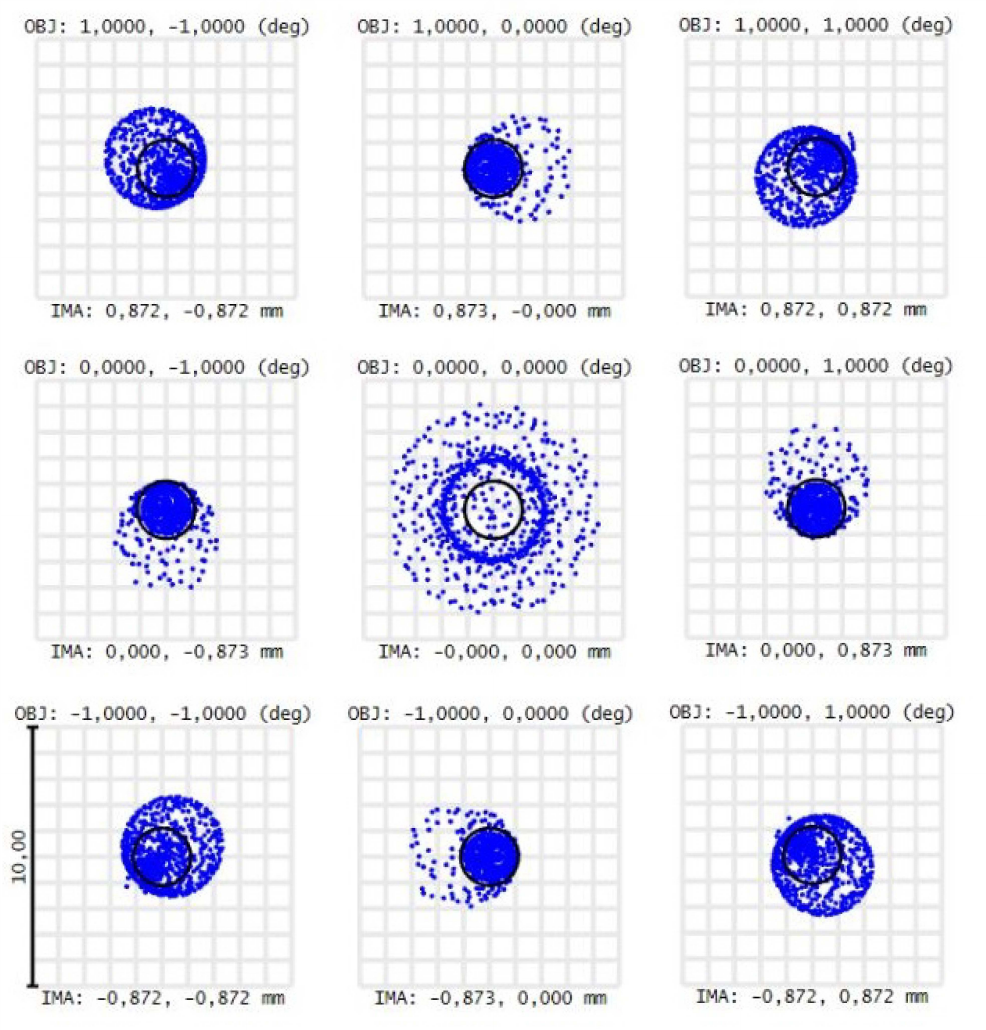}
			\end{tabular}
		\end{center}
		\caption 
		{ \label{fig:16}
			Spots in the image plane after optimizing the freeform corrector plate.} 
	\end{figure}
	Detailed analysis for the field point $Y = 1^o, X = 0^o$, using the described method is presented in Fig. \ref{fig:13}. It shows that the final correction is obtained due to the compensation of the lower order Zernike aberrations of the mirror [Fig. \ref{fig:13}(c)] with the freeform plate [Fig. \ref{fig:13}(a)]. Whereas the higher order Zernike aberrations are diminished by establishing a balance between induced, transfer and intrinsic components in the total system [Fig. \ref{fig:13}(d)]. Interesting to note is how the wavefront changes upon propagation from rear plane surface of the corrector plate until the mirror [Fig. \ref{fig:13}(b)].
	
	\begin{figure}[h]
		\begin{center}
			\begin{tabular}{c}
				\includegraphics[width=\linewidth]{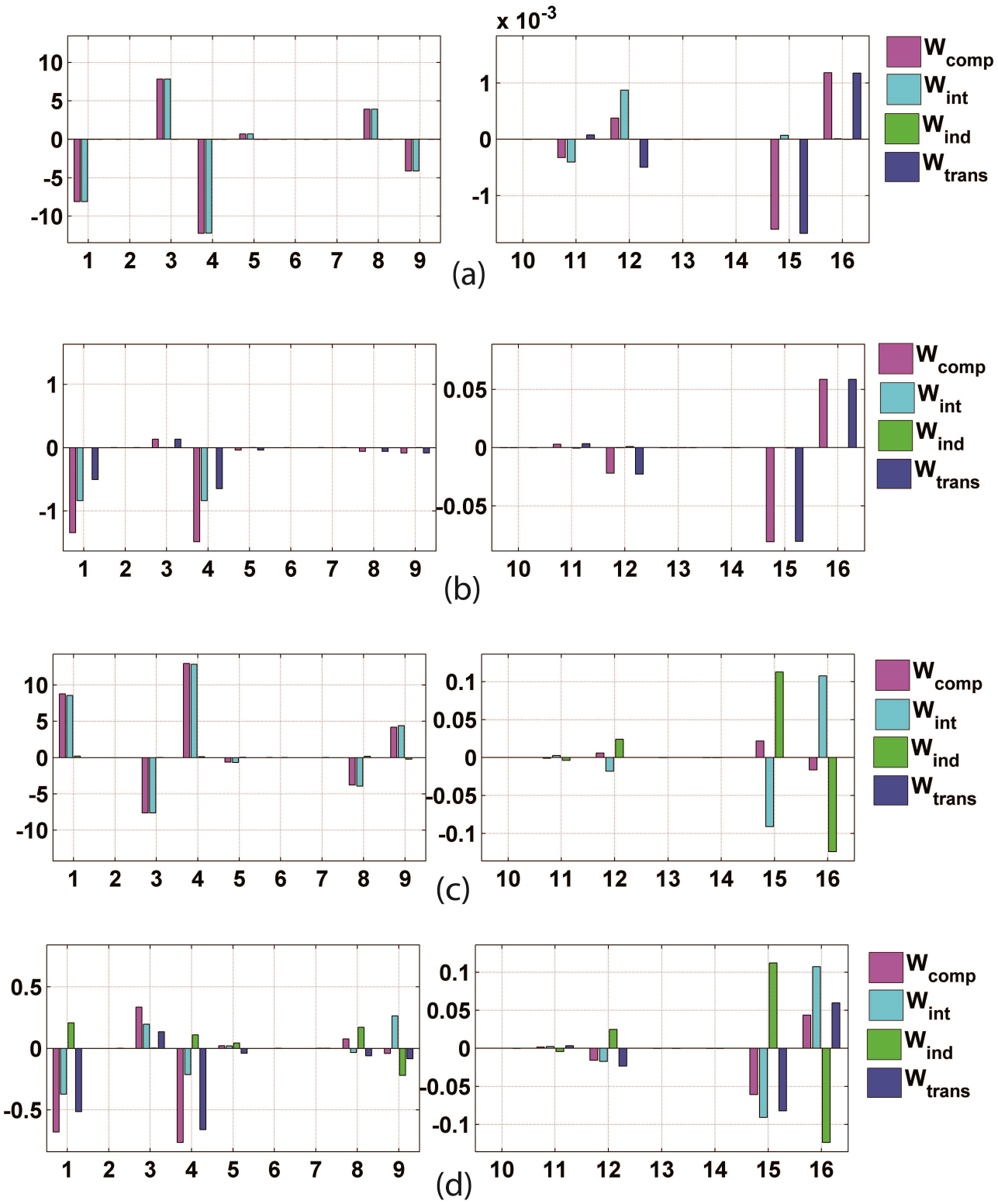}
			\end{tabular}
		\end{center}
		\caption 
		{ \label{fig:17}
			Zernike aberrations of respective surfaces shown with two axes for lower order up to Zernike 9 (left) and for higher order up to Zernike 16 (right). (a) Zernike aberrations of first, freeform surface, (b) Zernike aberrations of second, plane surface, (c) Zernike aberrations of third, mirror surface, (d) Zernike aberrations sum from all surfaces.} 
	\end{figure}
	The aberrations of the system can also be investigated for their field dependency using full field displays. Here two Zernike pairs, namely, Zernike coma (Z7 and Z8) and Zernike secondary coma (Z14 and Z15) are studied [Fig. \ref{fig:14}]. Two possibilities to represent the data graphically are shown. First, is to consider the complete contributions from respective surfaces [Figs. \ref{fig:14}(a) and (c)]. Second case is to examine the total aberration in the image plane with distinction into component types [Figs. \ref{fig:14}(b) and (d)]. As presented, depending on which Zernike aberration is studied, different configurations are more suitable. For lower order Zernike coma, insights are provided by looking at the complete surface contributions, whereas correction of Zernike secondary coma results from balance between induced, intrinsic and transfer components.\\
	\begin{figure}[h]
		\begin{center}
			\begin{tabular}{c}
				\includegraphics[width=\linewidth]{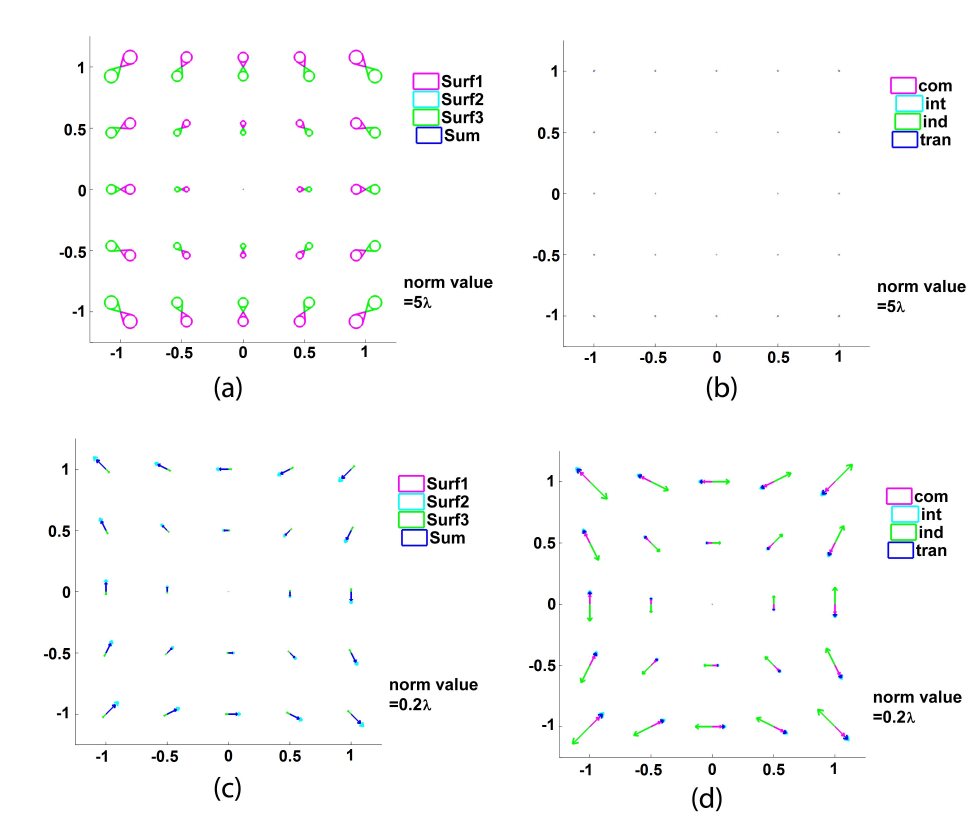}
			\end{tabular}
		\end{center}
		\caption 
		{ \label{fig:18}
			Full field displays of Zernike aberrations, (a) Zernike coma surface-by-surface contributions, (b) Zernike coma components of the total aberration (c) Zernike secondary coma surface-by-surface contributions, (d) Zernike secondary coma components of the total aberration.} 
	\end{figure}
	
	\section{Conclusion}
	\label{sec:7}
	We have presented the numerical approach to determine surface contributions to the total wave aberration in optical systems with no symmetry constraints. Surface contributions are defined as the wavefront change between entrance reference spheres located at the successive surfaces. The transfer component and the effect of refraction on the surface are treated separately as they correspond to independent design variables. Further refraction terms are distinguished due to their phenomenological origin into intrinsic and induced terms. The aberration of the chief ray is not included in the wave aberration and is considered separately, as the transverse error on the image plane. Due to those differences, the obtained results cannot be directly referred to the wave aberration coefficients. It is only possible for the special system presented in section \ref{sec:4}. \\
	The described method neglects vignetting and other aperture shape changes, assigning surfaces contributions to the uniform aperture shapes. For near-circular apertures it can be directly combined with the Zernike Fringe decomposition procedure.  This allows to perform detailed analysis of system Zernike aberrations with respect to individual surfaces and the component types. In section \ref{sec:6} an example is presented demonstrating that each of those divisions is suitable, depending on the considered aberration order. For lower order Zernike aberrations insights are provided by analyzing the compensation effect of individual complete surface contributions.  In case of coefficients higher than Zernike 9, the final correction is obtained due to the balance between respective components. \\
	The method has a potential to help selecting variables in the optimization process of systems with freeform elements.  It can be also utilized for finding critical differences between starting system configurations. Another possible application is tolerance sensitivity analysis.\\
	
	\section*{Acknowledgements}
	The authors want to thank Christoph Bösel for comments on the manuscript and for the grammar and spelling check.
	
	\section*{Funding}
	Federal Ministry of Education and Research Germany (fo+ - FKZ: O3WKCK1D)
	
	\section*{Acknowledgments}
	
	The authors want to thank Christoph Bösel for comments on the manuscript and for the grammar and spelling check.
	
	

\end{document}